\documentclass[sigconf,letterpaper,nonacm]{acmart}

\newcommand{\para }[1]{\medskip \noindent  {\bf #1}}
\newcommand{\1}{{\em (i)}}
\newcommand{\2}{{\em (ii)}}
\newcommand{\3}{{\em (iii)}}

\usepackage{xspace}
\newcommand{\sysnamelong}{DSP/RAM-Based Low-CLB Usage}
\newcommand{\sysname}{\textsc{DRAB-LOCUS}}
\newcommand{\thetitle}{\sysname{}: An Area-Efficient AES Architecture for Hardware Accelerator Co-Location on FPGAs}
\newcommand{\theshorttitle}{\sysname{}: Area-Efficient AES for Accelerator Co-Location on FPGAs}

\usepackage{booktabs}
\newcommand{\ra}[1]{\renewcommand{\arraystretch}{#1}}

\begin{document}

\title[\theshorttitle]{\thetitle}

\author{Jacob T. Grycel}
\email{jtgrycel@wpi.edu}
\affiliation{\institution{Worcester Polytechnic Institute}
  \department{Department of Computer Science}
  \city{Worcester}
  \state{MA}
}

\author{Robert J. Walls}
\email{rjwalls@wpi.edu}
\affiliation{\institution{Worcester Polytechnic Institute}
  \department{Department of Computer Science}
  \city{Worcester}
  \state{MA}
}

\ccsdesc[500]{Hardware~Hardware accelerators}
\ccsdesc[300]{Hardware~High-speed input / output}

\keywords{FPGA, AES, Low-Area, Throughput, Pipelined}

\begin{abstract}
Advanced Encryption Standard (AES) implementations on Field Programmable Gate 
  Arrays (FPGA) commonly focus on maximizing throughput at the cost of utilizing
  high volumes of FPGA slice logic. High resource usage limits systems'
  abilities to implement other functions (such as video processing or machine
  learning) that may want to share the same FPGA resources. In this paper, we
  address the \emph{shared resource challenge} by proposing and evaluating a
  \emph{low-area}, but high-throughput, AES architecture.
In contrast to existing work, our \sysnamelong{} (\sysname{}) architecture
  leverages block RAM tiles and Digital Signal Processing (DSP) slices to
  implement the AES Sub Bytes, Mix Columns, and Add Round Key sub-round
  transformations, reducing resource usage by a factor of 3 over traditional
  approaches.
To achieve area-efficiency, we built an inner-pipelined architecture using the
  internal registers of block RAM tiles and DSP slices. Our \sysname{}
  architecture features a 12-stage pipeline capable of producing 7.055 Gbps of
  interleaved encrypted or decrypted data, and only uses 909 Look Up tables,
  593 Flip Flops, 16 block RAMs, and 18 DSP slices in the target device.
\end{abstract} 
\maketitle
\section{Introduction}
\label{sec:intro}

As the Advanced Encryption Standard (AES) lies at the core of many important
security operations---ranging from securing network connections to full disk
encryption---improvements to performance and efficiency can benefit a large and
diverse set of systems. One means to achieve this performance increase is
through the use of \emph{hardware acceleration}. For instance, a number of
studies have leveraged field programmable gate arrays (FPGAs) for accelerating
AES~\cite{bib:tech-survey}.\footnote{While application-specific integrated
circuits (ASICs) offer an alternative approach to hardware acceleration, their
high non-recurring engineering cost and limited flexibility makes them less
attractive than FPGAs for use in low-cost systems.} While these FPGA-based AES
architectures achieve high throughput, they typically do so at the cost of high
resource usage on the FPGA, i.e., monopolizing large quantities of components
such as flip flops and look up tables.  While higher throughput is valuable, we
argue in this paper that \emph{area-efficiency} is often an equally important
design goal for AES architectures.

Area-efficiency, intuitively, is a measure of how effectively the AES
architecture capitalizes on the FPGA's resources. The key challenge is not only
making effective use of under-utilized components (such as digital signal
processing slices), but understanding the physical layout of those components
and incorporating that knowledge into the AES architecture. However, the
potential payoff is great as area efficient  designs offer substantial
benefits over those focusing purely on throughput.  Most notably,
area-efficiency opens up possibilities for using hardware-accelerated security
in new domains.  A variety of embedded systems simply cannot implement strong,
efficient encryption due to processor limitations or other constraints (e.g.,
power)---FPGA-based approaches are a promising way to overcome these challenges
and enable cryptographic security primitives on constrained embedded systems.  Indeed,
manufacturers have begun releasing cheap system-on-a-chip (SoC) platforms that
feature co-located FPGAs and CPUs. However, without area-efficient AES designs,
the system developer may have to choose between security and other operations
that benefit from hardware acceleration, such as deep
learning~\cite{bib:app-dlau, bib:app-dnn} or video
processing~\cite{bib:app-video}. An area-efficient AES architecture would allow
other types of hardware acceleration to run concurrently on the same FPGA.

In this paper, we propose a novel AES architecture that considers
resource-efficiency as a first-order design principle, balancing resource usage
and throughput.  Key to our design is the use of block RAM and digital signal
processing slices---resources that are largely ignored (or under-utilized) by
prior works~\cite{bib:tech-survey}---to efficiently implement the AES sub-round
transformations without the need for large numbers of logic slices. The
\sysnamelong{} (\sysname{}) architecture offers several advantages over
existing approaches, including: \1 high throughput on cheaper hardware; \2 more
efficient use of FPGA resources, including those left unused by most existing
AES architectures; and \3  more functionality, allowing for concurrent
encryption and decryption on multiple blocks.  We summarize our contributions
as follows:

\begin{itemize}

  \item We present the resource-efficient \sysname{} AES architecture which uses
    just 593 flip flops and 909 look up tables. The design leverages block RAM
    and digital signal processing resources to implement the sub-round
    transformations and build a 12 stage pipeline. We use the pipeline to
    construct an iterative and inner-pipelined datapath that is able to process
    12 independent blocks of data at any time. Furthermore, we use this
    architecture to produce 7.055 Gbps of data and can arbitrarily switch
    between encryption and decryption for any block in the datapath.

  \item We provide a deeper understanding of different AES architectures,
    exploring the fundamental trade-offs between throughput, resource usage,
    and power consumption. For instance, increasing the number of stages in a
    pipeline without increasing resource usage yields high resource-efficiency.
    These findings have implications beyond AES and can inform the design of
    resource-efficient architectures for other algorithms, both cryptographic
    and otherwise.

  \item We propose new metrics for evaluating implementation efficiency in
    Section~\ref{sec:efficiency} that incorporate other resource types like
    block RAM and digital signal processing resources.  For example, we consider
    throughput per look up table and throughput per flip flop, when before,
    previous studies consider only throughput per slice. These metrics provide a
    more complete evaluation of how effectively an implementation uses its
    resources to achieve high speed data processing, which can help designers to
    make informed decisions about which implementations to include in their
    system.

  \item We present a case study in co-location of hardware accelerators with
    AES implementations in embedded systems. This investigation explores what
    types of implementations may be more appropriate alongside different deep
    learning and video processing applications. We analyze whether using
    fully-unrolled AES architectures is feasible when resources are shared with
    other accelerators, and discuss the security benefits of area-efficient
    implementations in resource-constrained environments.

\end{itemize}

In Section~\ref{sec:background} we discuss pertinent background topics for AES
and FPGAs and then present our methodology for creating a high-speed and
area-efficient AES architecture in Section~\ref{sec:design}. We present our
\sysname{} implementation in Section~\ref{sec:implementation} and evaluate its
performance and efficiency compared to other recent AES architectures in
Section~\ref{sec:evaluation}. Additionally in Section~\ref{sec:evaluation}, we
recommend new metrics and standards for reporting and analyzing AES
implementations. We present our co-location case study considering AES
implementations and other hardware accelerators in
Section~\ref{sec:accelerators}, and provide a brief analysis of related
non-area-efficient work in Section~\ref{sec:related-work}. \section{Background}
\label{sec:background}

This section introduces the basics of the Advanced Encryption Standard and the
structure and components of FPGAs. The discussion of the \sysname{} design in
Section~\ref{sec:design} heavily references the topics introduced in this
section.

\subsection{The Advanced Encryption Standard}
\label{sec:aes}

The block cipher Rijndael is the base algorithm for the National Institute of
Standards and Technology's Advanced Encryption Standard (AES). Selected for its
simple architecture and easy implementation in both hardware and software, many
studies have focused on designing efficient digital circuit realizations of AES
and implementing them in Field Programmable Gate Arrays
(FPGA)~\cite{bib:tech-survey}. 

\subsubsection{AES Structure}
\label{sec:aes-struct}

AES is a symmetric-key block cipher with support for 128, 192, and 256-bit
keys. The cipher utilizes four \emph{sub-round transformations} and a \emph{key
schedule}, and treats a 128-bit \emph{input block} as a 4 byte by 4 byte
two-dimensional array called the \emph{cipher state}. The cipher state passes
through the sub-round transformations 10, 12 or 14 times (rounds) based on the
key size~\cite{bib:aes-spec}. We list the sub-round transformations and key
schedule below.

\begin{enumerate}
  \item \textbf{Sub Bytes:} A non-linear transformation performed by
    replacing an input byte based on a fixed look up table. The non-linear
    substitutions can also be calculated at runtime.
  \item \textbf{Shift Rows:}  A linear transformation on the rows of the state
    block, where each row is rotated a fixed number of positions.
  \item \textbf{Mix Columns:}  A linear transformation performed by
    multiplying the state block by a constant block using standard matrix
    multiplication.
  \item \textbf{Add Round Key:} A linear transformation that consists of
    adding the state block to a round key derived from the initial cipher key.
  \item \textbf{Key Schedule:} An iterative process that expands the initial
    128-bit key into 10 round keys by substituting bytes and rotating 32-bit words
    from the original and subsequent keys.
\end{enumerate}

\subsubsection{AES Mathematics in Hardware}
\label{sec:aes-math}

Mathematical operations in AES are performed on individual 8-bit values in
$GF(2^{8})$, a common algebraic finite field used in computer-based arithmetic.
Elements in this field are commonly represented as polynomials of degree 7 with
binary coefficients. Essentially, this is the finite field of two elements
extended to contain 8-term polynomials. By definition, both addition and
multiplication, as well as their inverses, can be performed on any element in
the field.

When represented as 8-bit binary strings, addition can be performed by
calculating the exclusive or (XOR) of two elements. Multiplication is slightly
more complicated, and can be implemented using a binary left shift followed by
an XOR reduction with a primitive polynomial that generates $GF(2^{8})$. These
simple operations in the field lend themselves to easy implementation in
digital circuits, which commonly contain both XOR and shifting components.
This close relationship between the cipher mathematics and digital hardware
resources makes AES an attractive cryptographic primitive for implementation in
digital systems.

\subsection{Field Programmable Gate Arrays}
\label{sec:fpga}

Field Programmable Gate Arrays (FPGA) are reconfigurable digital logic devices
that comprise combinatorial and sequential logic elements with a programmable
interconnect that allows digital design engineers to rapidly prototype systems
and update them in the field. Unlike Application-Specific Integrated Circuits,
FPGAs do not require long and expensive fabrication periods and are useful for
frequent hardware changes. As FPGAs are often used alongside a traditional CPU
to provide optimized hardware processing, there are more devices on the market
today that include CPUs and FPGAs into a single system on a chip to provide
tight integration of software and custom hardware. Because of this, FPGAs are
popular platforms for implementing hardware accelerators for applications such
as cryptography, deep learning, and video processing. Using hardware
accelerators can reduce CPU overhead by offloading processing-intensive tasks.

Below we discuss the primary components of the Xilinx 7-series FPGAs that
we utilize in our \sysname{} architecture: configurable logic blocks, block
RAM, digital signal processing slices, and clock regions.  We also discuss
finite state machines, a common control scheme for managing dataflow through
FPGAs. For a more complete introduction to FPGAs, see the survey by Kuon
et al~\cite{bib:fpga}.  

\subsubsection{Configurable Logic Blocks}
\label{sec:clb}

The majority of the FPGA consists of \emph{look up tables} (LUTs) and
\emph{flip flops} combined into units called \emph{configurable logic blocks}
(CLBs). LUTs implement combinatorial logic functions such as \texttt{AND},
\texttt{OR}, \texttt{XOR}, and basic math functions, where the result of the
function is stored for all possible combinations of inputs. Flip flops are
sequential storage units that run on a clock. On each edge of the clock, flip
flops store data on the input, and hold it on the output until the next clock
edge~\cite{bib:clb}.

LUTs offer two primary advantages over using individual logic gates. First, they
are able to implement large functions in a single component that would normally
require a high number of gates. This allows the FPGA to have a compact
architecture that increases its capacity for logic functions. Second, the delay
through the longest sequence of logic elements between two flip flops is easier
to predict. This sequence is also called the \emph{critical path}. With
logic gates, each component can have a different time delay from
when an input arrives and to when the gate produces an output. These varying
delays  make it difficult to calculate the total delay of the critical path.
With LUTs, each component has a nearly identical delay, so the critical path delay  
is determined by simply multiplying the LUT delay by the number of
path components.

Configurable logic blocks are organized in columns and rows within the FPGA to
simplify the connections made between each logic element and achieve high-speed
data rates. Each logic block is further divided into two logic slices that
contain multiple LUTs and flip flops. Refer to Appendix~\ref{apdx:fpga} for
further discussion of the layout of FPGAs and logic slices. As the most
abundant components of the FPGA, logic blocks have been highly optimized for
speed and flexible routing. For designs that primarily use slice logic, FPGA
design tools have higher flexibility for choosing locations at which to
implement logic functions. However, usage of other resources such as I/O ports,
clock buffers, processor interfaces, block RAM, and digital signal processing
slices (discussed next) can influence where designs place logic in the FPGA.

\subsubsection{Block RAM}
\label{sec:bram}

The second common resource in Xilinx 7-series FPGAs are block RAM tiles, which
are 36 kilobit memories accessible from within the FPGA\@. Block RAMs support
true dual-port access models, where each port (a or b) can run on an
independent clocks and have separate addresses and write enables.  Our
architecture uses  the dual-port functionality to utilize a single block ram to
perform look up operations on two bytes of the cipher state, as discussed in
Sections~\ref{sec:sb} and~\ref{sec:mc}.

Xilinx block RAM supports data widths up to 32 bits wide, and can be divided
into any number of storage locations, as long as the number of data address can
be encoded in 16 bits. Depending on the size of the needed memory, block RAM
tiles may be configured as two separate 18 kilobit blocks, or a single 36 Kb
block. Refer to Appendix~\ref{apdx:fpga} for further discussion of block RAM\@.

Block RAM memory look ups have a latency of one cycle by default, however they
also contain a built-in register that can delay output data by an extra clock
cycle. This is useful because the block RAM devices write output data late in
the clock cycle, which makes it difficult for the output to reach other
sequential elements in time for the next rising edge. By enabling the output
register, outputs from block RAM will appear soon after the rising edge, and
have more time to travel through the critical path and reach the next sequential
element.

\subsubsection{Digital Signal Processing Slices}
\label{sec:dsp}

The last relevant Xilinx 7-series resources are the Digital Signal Processing
(DSP) slices. DSP slices were introduced to FPGAs to increase signal processing
capabilities for digital and analog signal applications. These blocks can
implement a number of mathematical and logical operations on 48-bit wide
inputs, and are organized in the FPGA for high speed daisy-chaining along
vertical paths in the fabric. Optionally, input data can be delayed by 1 or 2
clock cycles using the internal registers for implementing digital signal
filters, or for decreasing delays between connected components for meeting
timing constraints~\cite{bib:dsp}. We leverage the internal registers to build
pipeline stages into our design without having to use lots of flip flops from
logic slices, as discussed in Sections~\ref{sec:sb},~\ref{sec:mc},
and~\ref{sec:ark}. See Appendix~\ref{apdx:fpga} for more details on DSP
resources.

\subsubsection{Clock Regions}
\label{sec:clock-region}

FPGAs are divided into partitions called \emph{clock regions}, which consist of
columns of logic blocks, block RAM, and DSP slices. While each clock region has access
to global clocks that reach every section of the FPGA, they also contain
dedicated paths for local clock signals not accessible anywhere else.
The organization of FPGAs into clock regions allows for more complicated
designs that require  multiple distinct operating
frequencies. One further advantage is that if an FPGA is running a design that
only occupies a single clock region, it is possible to constrain the clocks to only
run in that region and avoid wasting power on sending the signals to other parts
of the device~\cite{bib:clock}.

\subsection{Finite State Machines}
\label{sec:fsm}

FPGA designs often require a control module that is responsible for directing
data, and configuring other parts of the design to perform different actions.
One of the most common techniques for implementing controllers is to create a
finite state machine that modifies outputs based on inputs from the system. 
Finite state machines are easily implemented in FPGAs, with flip flops that hold
the current state of the controller, and LUTs that calculate the next state to
transition to based on the state flip flops and any inputs to the design. Due to
the compact organization of LUTs and flip flops in the logic slices, state
machines are usually implemented in such a way that they are able to run at any
speed supported by other components in the FPGA\@.
 \section{Area-efficient Design}
\label{sec:design}

We developed our FPGA-based \sysname{} AES architecture by considering
alternative resource usage strategies. In order to develop an area-efficient
and high-throughput implementation, we considered a number of different
high-level architectures and sub-round transformation designs. Whereas other
studies tend to focus on optimizing a single section of the AES algorithm, we
set out to design a full architecture and consider how low-level optimizations
and design techniques would impact the cipher as a whole. Throughout the design
process, we used the following performance indicators to gauge the effect our
design decisions would have on the overall architecture: \1 \emph{throughput},
the number of encrypted/decrypted bits produced each second, \2 \emph{Latency},
the number of cycles elapsed before a block finishes all AES rounds, \3
\emph{resource usage},  the number of flip flop, look up table, block RAM, and
DSP slices used across the FPGA\@.

There are a number of FPGA design techniques that can affect these performance
indicators. For example, running the design on a faster clock will increase the
throughput of the design, but make it harder to use alternative resources like
block RAMs and DSP slices since they are placed farther apart. Logic slices are
better suited for high speed operation, which means more and more parts of the
design will need to run in logic slices as the clock speed increases. However,
placing flip flops in between block RAM and DSP slices can reduce the critical
path between these distantly placed components, making it easier to use them
and run at a high clock rate. An example of this is discussed in
Section~\ref{sec:sr}. 

The remainder of this section describes the full \sysname{} design. We compare
our design to other architectures in Section~\ref{sec:evaluation}.  We leave
the exploration of possible  side channels and physical layer attacks for future
work and instead focus on the performance and architectural advantages of our
design. 

\subsection{Datapath Architecture}
\label{sec:datapath}

A range of architectures have been used in the past to implement
AES in hardware, each of which can be defined by whether it uses pipelining or
loop-unrolling techniques. In a pipelined design, registers are inserted into
the datapath at regular intervals to increase the amount of data that can be
processed at one time. Furthermore, registers can be inserted inside of a single
AES round, or between subsequent copies of the round, given that the design uses
multiple round instances; these two techniques are referred to as inner-round
and outer-round pipelining, respectively~\cite{bib:tech-survey}.

In a loop-unrolled design, an entire round is instantiated multiple times with
data flowing sequentially through each copy, allowing for data to continuously
enter the cipher without any control mechanism. On the other hand, in an
iterative architecture, data is fed back into a single instance of the round
multiple times. When paired with inner and/or outer-round pipelining, designs
can produce high throughput at the expense of more resource usage (more pipeline
stages yield more flip flops).

\begin{figure*}[t]
  \centering
  \includegraphics[width=\textwidth,keepaspectratio]{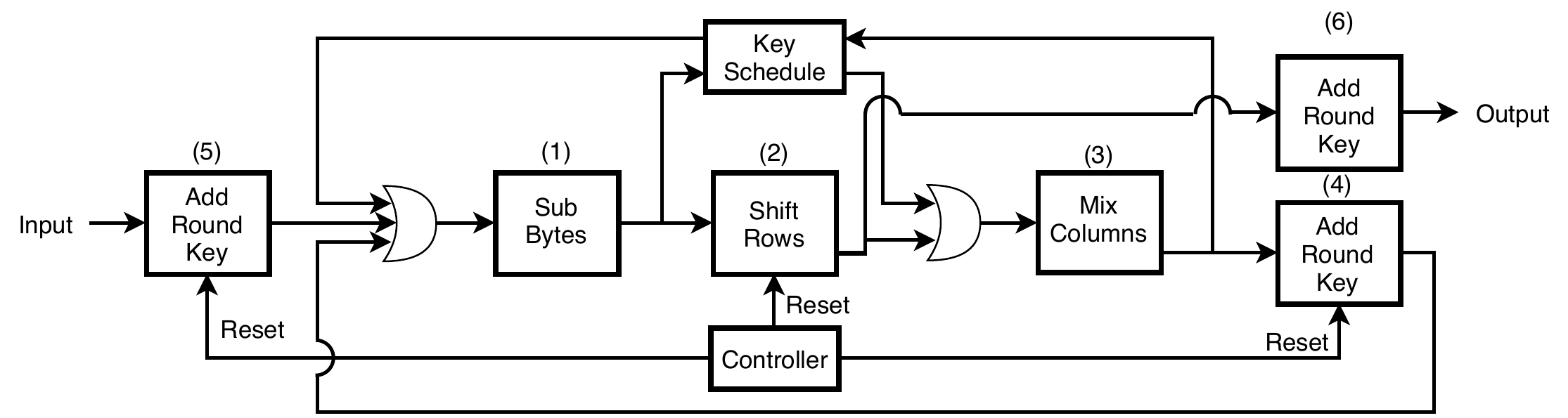}
  \caption{Full Datapath Architecture}
  \label{fig:dp-full} 
\end{figure*}

We designed the \sysname{} architecture with an iterative, inner-pipelined
structure, allowing us to maximize throughput while using few resources. This
means the datapath mainly consists of a single instance of each sub-round
transformation, where data is fed back through the design until it has completed
9 rounds, as can be seen in Figure~\ref{fig:dp-full}. While the iterative
component of our design minimizes resource usage by only instantiating the AES
round once, the inner-pipelined aspect requires more hardware registers, which
can be expensive depending on the number of pipeline stages. In order to
mitigate this, we use the built-in registers of the block RAM and DSP slices to
insert pipeline stages, as explained in subsequent sections. All register stages
are indicated by red dashed boxes in the datapath figures.

We modify the iterative inner-pipelined structure to include two extra instances
of the add round key transformation (5,6) in order to easily handle multiple
blocks of data at the same time. With a single instance of the add round key
transformation (4), there could be a case where two blocks need to pass through
the same instance at the same time. For example, if one block exits the mix
columns transformation (3), and one block exits the shift rows transformation
(2) during its final round, both blocks have to pass through the add round key
transformation (4) next. With a single instance, a controller would have to
stall the entire datapath to resolve the data conflict. By using three instances
of the add round key transformation (4,5,6), we avoid potential data hazards.

We were able to use this same architecture for both encryption and
decryption by using the \emph{equivalent inverse cipher}~\cite{bib:aes-spec}.
This is a modification to the AES decryption algorithm that reorders the inverse
sub-round transformations to be in the same order as their encryption
counterparts, with the only significant change being required in the key
schedule. In the equivalent inverse cipher, each round key must pass through the
inverse mix columns transformation (3) before being added to the cipher state in
the add round key transformation (4,6). To support this, we added connections to
the datapath before and after the mix columns transformation (3) for the key
schedule to perform inverse mix columns on each round key. Additionally, we
added connections before and after the sub bytes transformation (1) to perform
s-box lookups during round key calculation, as discussed in
Section~\ref{sec:ks}.

As the \sysname{} architecture supports arbitrary switching between encryption
and decryption at every pipeline stage, the key schedule must calculate all
round keys and inverse round keys during an initialization phase, so during
cipher operation the key schedule taps into the datpath are unused
(Section~\ref{sec:ks}). This means that we can use an \texttt{OR} gate to
multiplex sub-round transformations, as the key schedule inputs are held at
zero. This reduces the complexity and connectivity of the controller, making the
design easier to implement. The only other requirement for using \texttt{OR}
gates is that the output of the add round key transformation (4) is reset to
zero when a new input is added to the datapath. The controller triggers this
reset using dedicated high-speed reset paths in the FPGA, as discussed in
Section~\ref{sec:control}. A high-level block diagram of the data path
with control and key schedule connections is shown in
Figure~\ref{fig:dp-full} and a  schematic of the datapath can be seen
in Figure~\ref{fig:arch}.

\subsubsection{The Sub Bytes Transformation}
\label{sec:sb}

The \sysname{} architecture performs the sub bytes transformation by using block
RAM tiles configured as ROM look up tables. This is one of two commonly used
techniques, the other of which is to perform composite field arithmetic in
$GF(2^{8})$~\cite{bib:tech-survey}. We chose to use the block RAM look up table
technique as this results in less utilization of slice logic.

The sub bytes design uses each byte of the input block as an address into a
block RAM look up table. There is enough space to store the tables for both
encryption and decryption such that the encryption table resides in the first
2,048 bits of memory, and the decryption table resides in the second 2,048 bits
of memory. Therefore, we use a single block RAM to perform both the sub bytes
and inverse sub bytes transformations, by prepending the mode for the current
block to each 8-bit address input to select between the encryption and
decryption tables.

\begin{figure}[t]
  \centering
  \includegraphics[width=0.5\textwidth,keepaspectratio]{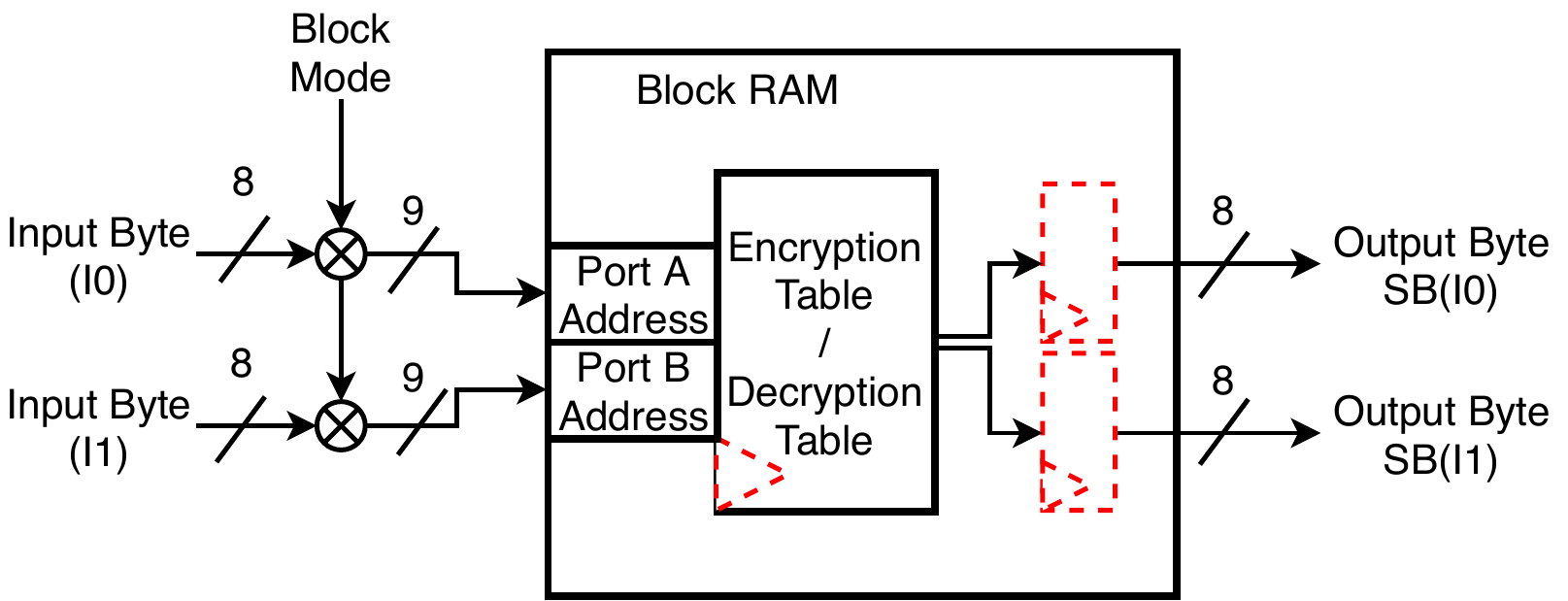}
  \caption{Sub Bytes Transformation Design}
  \label{fig:sub-bytes}
\end{figure}

Furthermore, since the Xilinx 7-series block RAM supports true dual port
interfacing, we use a single block RAM to perform lookups for two bytes of the
input block, as shown in Figure~\ref{fig:sub-bytes}. To construct the full sub
bytes transformation, we replicate this structure 8 times in the full datapath.
Although we strive to minimize resource usage, replicating the block RAM 8 times
greatly simplifies the \sysname{} architecture, as using a single block RAM
would add internal loops to the datapath. This would require additional control
logic and make it difficult to process multiple blocks with different modes at
the same time.

The sub bytes transformation adds two pipeline stages to the datapath, where the
first stage comes from the 1-cycle memory look up, and the second from enabling
the output registers built into the block RAM\@. The \sysname{} architecture
benefits from the second stage in two ways. First, the lookup delay through
the RAM is slow, and can make it hard to implement the design in the FPGA
without causing timing issues. The output register makes it easier to implement
the design with a fast clock (as discussed in Section~\ref{sec:bram}). Second,
as the second register is internal to the block RAM, this increases the number
of pipeline stages in the datapath without requiring 128 flip flops from the logic
slices. Both of these features contribute to high throughput
and low resource usage.

\subsubsection{The Shift Rows Transformation}
\label{sec:sr}

The design implements the shift rows transformation using LUTs and flip flops
from the logic slices. Assuming that we only wanted to support encryption (or
decryption), we could statically maps input bytes to rotated output bytes.
However, because the \sysname{} architecture concurrently performs both
encryption and decryption, the datapath uses a switch made of LUTs and flip
flops to choose between the rotations for encryption or decryption depending on
the mode needed for the current block.

While using logic slices to implement shift rows requires the use of 128 flip
flops, this technique also allows the cipher to run at higher clock frequencies,
as the flip flops can be placed equidistant to the sub bytes and mix columns
instances. This decreases the critical path between block RAMs, which may be
placed far apart due to the layout of Xilinx 7-series FPGAs, allowing for an
overall faster cipher speed.

\subsubsection{The Mix Columns Transformation}
\label{sec:mc}

\sysname{} splits the mix columns transformation into two sequential stages,
corresponding to the two steps of matrix multiplication.

\textbf{Byte Multiplication Lookup} -- The first step of matrix multiplication
is to multiply elements of the two matrices together. We load precomputed byte
multiplications corresponding to the mix columns matrix into block RAMs, which
perform a lookup operation for each byte of the input block. Since the mix
columns input block is multiplied by a fixed matrix for encryption and
decryption, it is guaranteed that each byte of the input block is multiplied by
every byte of the fixed matrix at some point during the computation.
Furthermore, as each row of the fixed matrix is simply a rotational permutation
of the first row, each input byte will only be multiplied by four different
values. For example, each input byte is multiplied only by 01 (twice), 02, and
03 for encryption; for decryption, each byte is multiplied by 09, 0B, 0D, and
0E. Additionally, both matrices have the same structure, but with different
bytes. This allows us to use the same storage structure in block RAM for
encryption and decryption. We discuss the storage format further in
Appendix~\ref{apdx:mc}. Each input block byte is used as an address to look up
the corresponding 32-bit multiplication results, and the mode
(encryption/decryption) is used to select the high or low 256 memory entries
(similar to the sub bytes transformation design).

We use the registers built into the block RAM to add another pipeline stage to
the datapath, again without using any flip flops. Similar to the sub bytes
design, this also enables the cipher to run at higher frequencies, as the
output register decreases the length of the critical path within the mix
columns instance.

\textbf{Wide XOR} -- The second step of matrix multiplication is to add all of
the multiplied terms together to produce an element in the output matrix. The
design uses DSP slices to perform 48-bit wide \texttt{XORs} on the outputs of
the block RAM look up. We mix the block RAM outputs together to form 4 48-bit
vectors for the high 48 bits, 4 48-bit vectors for the middle 48 bits, and 4 
32-bit vectors for the remaining 32 low bits. Equation~\ref{eq:mc-mult} shows an
example of the standard matrix multiplication for the first 4 output bytes
during encryption, and we use the patterns of matrix multiplication to
concatenate different output bytes from the block RAMs, such that addend terms
line up in the same position across four vectors. Refer to
Appendix~\ref{apdx:mc} for more specific equations involving the block RAM
outputs. We assume that the input state is organized in column-major order as
shown in $S$, with $s_{0,0}$ containing the MSB\@.

\begin{center}
  $S = \quad
    \stackrel{\mbox{Cipher State Layout}}{\begin{tabular}{|c|c|c|c|}
      \hline
      $s_{0,0}$ & $s_{0,1}$ & $s_{0,2}$ & $s_{0,3}$ \\
      \hline
      $s_{1,0}$ & $s_{1,1}$ & $s_{1,2}$ & $s_{1,3}$ \\
      \hline
      $s_{2,0}$ & $s_{2,1}$ & $s_{2,2}$ & $s_{2,3}$ \\
      \hline
      $s_{3,0}$ & $s_{3,1}$ & $s_{3,2}$ & $s_{3,3}$ \\
      \hline
    \end{tabular}}
  $\end{center}

\begin{gather}
  \begin{split}
    Out_0 = 02*s_{0,0} + 03*s_{1,0} + 01*s_{2,0} + 01*s_{3,0} \\
    Out_1 = 01*s_{0,0} + 02*s_{1,0} + 03*s_{2,0} + 01*s_{3,0} \\
    Out_2 = 01*s_{0,0} + 01*s_{1,0} + 02*s_{2,0} + 03*s_{3,0} \\
    Out_3 = 03*s_{0,0} + 01*s_{1,0} + 01*s_{2,0} + 02*s_{3,0}
  \end{split}
\label{eq:mc-mult}
\end{gather}

The 48-bit vectors pass into cascaded DSP slices, such that the output of one
slice is the input to another. A cascaded structure improves the maximum
operating speed of the DSP slices by using dedicated short-distance routes
between the slices. To operate this part of mix columns at the same
speed as the rest of the architecture, we enabled internal DSP slice input and
output registers, decreasing the critical path between DSP slices and block
RAMs.

The number of registers used for each DSP slice varies in order to synchronize
data through the cascade. For example, the inputs to the second slice must be
delayed by two cycles to account for the 2-cycle delay through the first slice.
In order to use an input and output register in the first slice, a total of 
three delay cycles are required for the third slice in the cascade. Since the
DSP slices only have 2 input registers, mix columns uses 128 flip flops from the
logic slices to acquire the third delay cycle. In total, the DSP cascade tree
adds four pipeline stages to the datapath, in addition to the 2 stages from the
block RAMs. The full mix columns design is shown in the full datapath diagram
(Figure~\ref{fig:arch}).

\subsubsection{The Add Round Key Transformation}
\label{sec:ark}

The \sysname{} architecture implements the add round key transformation using
DSP slices, similar to the second stage of mix columns. Since the add round key
transformation \texttt{XORs} two 128-bit values, only three parallel DSP slices
are needed, unlike mix columns, which cascades sequential DSP slices. Since the
DSP slices operate on 48-bit signals and there is a 128-bit input, two DSP
slices compute two 48-bit \texttt{XORs} and one DSP slice computes one 32-bit
\texttt{XOR}. In order to add more pipeline stages to the datapath, and to
decrease the critical path between the add round key transformation and the
mix columns transformation and key schedule, the design enables two input
registers and one output register on each DSP slice.

For the two extra add round key instances in the initial and final rounds of the
cipher, the architecture only uses one input register, since the DSP slices are
only routed to the key schedule, and not to block RAMs (as in the main-round
instance). Since the initial and final add round key instances are not part of
the main datapath pipeline, extra registers do not contribute to throughput, and 
instead only increase the latency. Thus, by only using two register stages we
maintain a shorter critical path to the key schedule and reduce overall latency
by two clock cycles.

\begin{figure*}[h]
  \centering
  \includegraphics[width=0.8\textwidth,keepaspectratio]{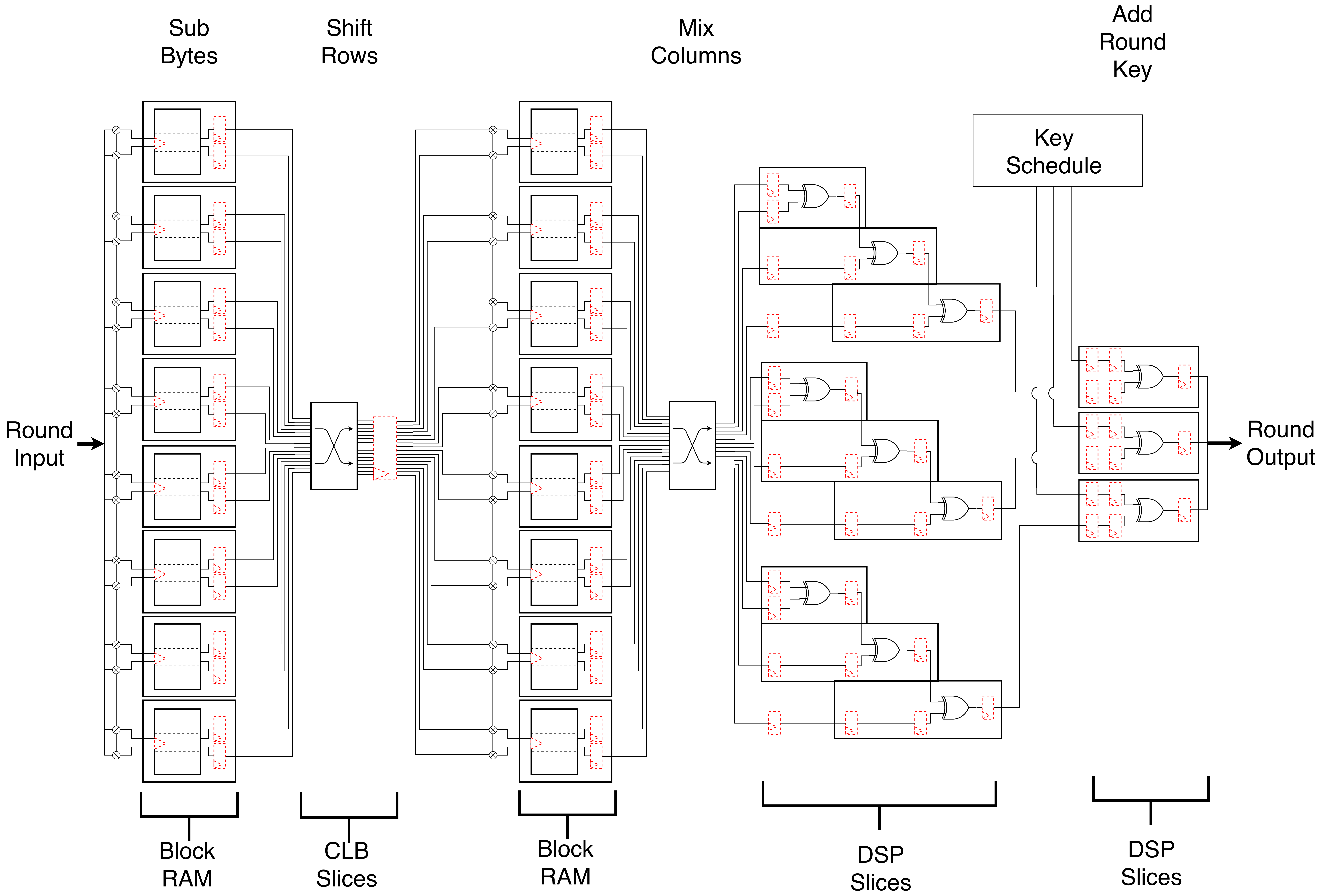}
  \caption{Full Datapath Schematic}
  \label{fig:arch}
\end{figure*}

One final component of the DSP slices we use are the reset inputs, which hold
the outputs at zero when a new block enters the pipeline, or the key schedule is
initializing the round keys. This reset signal, and the data hazards it
prevents, are discussed further in Section~\ref{sec:control}. The three parallel
DSP slices for the add round key instance can be seen in the datapath in
Figure~\ref{fig:arch}.

\subsection{Control Module}
\label{sec:control}

Since the datapath for our \sysname{} architecture has an iterative
inner-pipelined structure, we designed a controller to ensure that each block
passes through the transformations the appropriate number of rounds and to
prevent data collisions when a new block enters the datapath. Since the design
processes multiple blocks at the same time, new input blocks have to be
synchronized into the datapath so as to not interfere with data in the pipeline.
We designed the control module so it meets the following requirements:

\begin{enumerate}
  \item Track each block as it moves through the datapath until it is ready to
    pass through the final add round key instance.
  \item Pass the mode of operation (encryption/decryption) for each
    block to the sub-round transformation it is currently in.
  \item Stall new inputs to the cipher when data in the pipeline is moving from
    add round key to sub bytes.
  \item Reset the output of the initial add round key instance until a new block
    of data is ready to enter the pipeline.
  \item Hold the output of the add round key instance in reset when a new block
    of data enters the pipeline.
  \item Hold the add round key and shift rows instances in reset while the key
    schedule calculates round keys.
\end{enumerate}

The controller fulfills requirement 1 by using 12 113-bit-long shift registers
to track each block as it travels through each stage of the pipeline and
iterates through the datapath until the final add round key transformation. We
implement these using LUTs from the M-type logic slices, which feature LUTs with
clock inputs that can be used as 32-bit shift registers, as long as the only bit
that needs to be accessed is the final bit~\cite{bib:clb}. Since the \sysname{}
controller only needs the last bit from the 113-bit shift registers to know when
a block has completed the appropriate number of rounds, we can implement the
registers using 4 LUTs from the M-type logic slices without requiring any flip
flops.

The controller fulfills requirements 2, 3, 4, and 5 using two 12-bit-long shift
registers, where one tracks which pipeline stages have a block in them, and the
other tracks the mode for each block in the pipeline. The pipeline-tracking
shift register controls whether the add round and initial round key instances
should be reset (requirements 3, 4, and 5). The mode shift register is
responsible for passing the correct mode of operation to each sub-round
transformation so that the correct operation is performed on the block currently
in each pipeline stage (requirement 2). We implement these shift registers using
LUTs and flip flops, since the controller must access multiple bits inside each
register, making it impossible to use the clocked M-type LUTs.

The controller fulfills requirement 6 by using a finite state machine to
identify when the key schedule is initializing, or the entire cipher is reset.
During reset and initialization, the state machine holds data in the datapath at
a zero value, and releases it for normal operation. The state machine also has
a flush state that fills the shift registers with zeros, as the shift-register
configuration of the LUTs doesn't accept a reset signal.

\subsection{Key Schedule}
\label{sec:ks}

We designed the key schedule with a finite state machine that iteratively
computes round keys and inverse round keys for the equivalent inverse cipher so
that the key for any round is available on request since any block in the
pipeline can be in any round. We use block RAM to store all of the computed
keys, and use 12 counters to keep track of which round key is needed for each
block in the datapath. As with the sub bytes and mix columns block RAM, we
append the block mode bit to the memory look up address to select between keys
for encryption and decryption.

As described in Section~\ref{sec:datapath}, the key schedule logic connects to
the sub bytes and mix columns transformations, as round key calculations use
s-box substitutions, and the equivalent inverse cipher requires inverse mix
columns to each key for decryption. Since the controller holds the rest of the
datapath in reset during initialization, the key schedule is able to use these
transformations without data errors. During normal cipher operation, the key
schedule state machine holds its inputs to these transformations in reset to
avoid data errors as well.

Due to having three instances of the add round key transformation, it is
possible for the datapath to need the initial, final, and an arbitrary round key
at the same time. We solve this challenge by using one block RAM port to produce
an arbitrary round key and another port to always produce the final round key,
and by adding a 128-bit register to hold the initial round key. This register,
and two other 128-bit registers that hold intermediate keys during
initialization, make the key schedule have the most slice usage of the design.
This is reflected in Table~\ref{tab:full-usage} in the next section.

\para{Summary:} Our usage of block RAMs and DSPs allows us to build a 12-stage
pipeline for the \sysname{} datapath without relying on slice logic to
implement pipeline stages or sub-round transformations. We use the built in
registers of block RAM and DSPs to add stages to the pipeline. A small control
module accompanies the iterative and inner-pipelined datapath to ensure correct
dataflow through the cipher, and also keeps track of whether each block is
undergoing encryption or decryption. Finally, we take advantage of the
equivalent inverse cipher for AES to use the same datapath for both encryption
and decryption, and design the key schedule to be able to produce any round key
for any block in the datapath, regardless if using encryption or decryption.
 \section{Implementation}
\label{sec:implementation}

We synthesized and implemented the \sysname{} design using a high-speed-grade
Zynq 7000 SoC featuring co-located ARM processors and an Artix-7 grade FPGA
(xc7z030sbg485-3). As a result of our focus on using block RAM and DSP slices
for area-efficiency, the entire implementation fits in one half of a clock
region. This uses less power because the FPGA only has to route clock signals to
one section of the device. 

In addition to fitting  within a single clock region,  all of the logic
elements are contained within two columns of block RAM titles. This compact
layout allows \sysname{} to run at the maximum frequency supported by the block
RAMs. This is because there is less physical distance between components and,
therefore, less delay on the critical path.

This implementation of the \sysname{} architecture runs on a 528 MHz clock,
producing 7.055 Gbps of interleaved encrypted and decrypted data with a latency
of 217 nanoseconds.  Table~\ref{tab:full-usage} gives a detailed breakdown of
the resource and power usage. In Appendix~\ref{apdx:physical}, we provide more details on
the physical layout. We use Equations~\ref{eq:latency}
and~\ref{eq:throughput} to calculate latency and throughput, respectively, as
is done in other FPGA implementation studies~\cite{bib:ipit1,bib:cmp-paar}.

\begin{table*}\centering
  \caption{Resource usage and power consumption of \sysname{} components}
  \sffamily
  \ra{1.3}
  \begin{tabular}{@{}lrrrrrcrrrrr@{}}
    \toprule
    & \multicolumn{5}{c}{Resource Usage (\#)}
    & \phantom{abc}
    & \multicolumn{5}{c}{Power Consumption (mW)}\\
    \cmidrule{2-6}
    \cmidrule{8-12}
        &\multicolumn{1}{c}{\textbf{Slices}} & 
        \multicolumn{1}{c}{\textbf{LUTs}} & 
        \multicolumn{1}{c}{\textbf{Flip Flops}} & 
        \multicolumn{1}{c}{\textbf{B. RAMs}} & 
        \multicolumn{1}{c}{\textbf{DSPs}} &
        &
        \multicolumn{1}{c}{\textbf{Slices}} & 
        \multicolumn{1}{c}{\textbf{B. RAMs}} & 
        \multicolumn{1}{c}{\textbf{DSPs}} & 
        \multicolumn{1}{c}{\textbf{Signal}} & 
        \multicolumn{1}{c}{\textbf{Clock}} \\
    \midrule
    \textbf{Control}      & 21    & 27  & 34   & --    & -- && 0.5 & --     & --    & 0.5 & 3\\
    \textbf{Key Schedule} & 220   & 616 & 303  & 4     & -- && 8 & 99 & --  & 22 & 21\\
    \textbf{Datapath} \\
    \quad Sub Bytes       & 23    & 64  & --   & 4     & -- && 3 & 86  & --    & 12  & 0.5\\
    \quad Shift Rows            & 74    & 74  & 128  & --    & -- && 1   & --     & --      & 8   & 8\\
    \quad Mix Columns           & 38    & --  & 128  & 8     & 9 && 0.5 & 173  & 35   & 42  & 10\\
    \quad Add Round Key         & 15    & 32  & --   & --    & 3 && 0.5  & --     & 13   & 3   & 0.5\\
    \quad \textasciicircum Init  & 42    & 96  & --   & --    & 3 && 2    & --     & 0.5  & 3   & 0.5\\
    \quad \textasciicircum Final & --    & --  & --   & --    & 3 && --     & --     & 9    & --     & 0.5\\
    \addlinespace
    \textbf{Total (data)}     & 167   & 266 & 256  & 12    & 18 && 7 & 259 & 58 & 68 & 20 \\
    \textbf{Total (all)}    & 310   & 909 & 593  & 16    & 18  && 15.5 & 357 & 58 & 90.5 & 44 \\
    \bottomrule
  \end{tabular}
  \label{tab:full-usage}
\end{table*}

\begin{equation}
  \begin{split}
    Latency = ({12\,stages} * 9 + 7\,cycles) * {1.89\,ns \over 1\,cycle}
  \end{split}
\label{eq:latency}
\end{equation}

Latency is defined as the  number of cycles required to process a single
block of data. Each block passes through the 12-stage datapath pipeline for 9
rounds, and then through the sub bytes, shift rows, initial add round key, and
final add round key transformations to complete the initial and final rounds.
We multiply the cycle count by the clock period to obtain the latency in
nanoseconds.

\begin{equation}
  \begin{split}
    Throughput = {128\,bits \over 1\,block} * {528.262\,M\,cycles \over 1\,second} * {1\,block \over 115\,cycles} * {12\:blocks}
  \end{split}
\label{eq:throughput}
\end{equation}

The throughput is calculated as follows.  First, by dividing the clock
frequency by the cycle count  we know how often the design produces a
block of data. We multiply this rate by 128 to account for the 128 bits in each
block. Finally, we multiply  by 12 to account for the 12 blocks processed
concurrently in the datapath (12 blocks of data are produced every 115 cycles).

\para{Summary:} The \sysname{} implementation uses only 909 LUTs, 593 flip
flops, 16 block RAM, and 18 DSP in the target FPGA. This low resource
utilization produces a compact layout that allows the design to run on a fast
clock signal, where the clock speed is limited by the maximum block RAM
frequency  and not the critical path delay. The AES sub-round transformations
use only 266 LUTs and 256 flip flops, less than half of the total design's
slice utilization.
 \newcommand{\AESencdec}{AES-EncDec}
\newcommand{\AESmodes}{AES-Modes}
\newcommand{\AESpaar}{AES-Efficient}
\newcommand{\AESexpand}{AES-Expanded}

\section{Evaluation}
\label{sec:evaluation}

Our evaluation focuses on answering the following key questions.  First, how
does \sysname{} compare to other FPGA-based AES architectures? Specifically,
what is the impact of different  architectural design decisions on throughput
and resource usage?  Second, how efficiently does \sysname{} use its allocated
resources?  How do we evaluate efficiency for AES designs in general?  Finally,
how does the use of non-logic-slice resources affect power consumption? 

We pick the following three architectures for our evaluation because \1 they
all leverage block RAM and DSP resources and \2 they all optimize for different
metrics: \emph{\AESencdec{}}  optimizes for high-throughput, \emph{\AESmodes{}}
optimizes for low-area, and \emph{\AESpaar{}} optimizes for area-efficiency.
The latter architecture offers the most direct and interesting comparison to
\sysname{}.
 
As the Xilinx design tools compute the resource usage, power consumption, and
clock frequencies of the implementations, we did not have to load the designs
on to physical chips to measure performance. Instead we use the implementation
statistics from the studies, which we summarize in Table~\ref{tab:compare}.
Beyond these three designs, we provide a brief analysis of more AES
implementations in Section~\ref{sec:related-work} to compare \sysname{} to
designs that do not leverage block RAM and DSP, or do not describe how they use
them.

\para{\AESencdec{} architecture.} Wang and Ha achieved a throughput of
78.22 Gbps using a fully-pipelined architecture at the cost of high usage of
all types of FPGA components: flip flops, LUTs, Block RAM,
DSP~\cite{bib:cmp-encdec}. This is an example of a high-throughput design with
high resource usage. We also include it to discuss how architecture structure
can increase design performance at the expense of high resource usage, even when
two designs use the resource in the same way. Additionally, this is the only
comparable design that supports both encryption and decryption, as \sysname{}
does.

\para{\AESmodes{} architecture.} de la Piedra et al.\ built a low-area design
using 28 DSP slices, and only 159 logic slices and 15 block RAM. However, their
design has a much lower throughput than \sysname{} with only 124
Mbps~\cite{bib:cmp-modes}. This is an example of a design with low resource
usage that produces low throughput. We also include it to discuss how
architecture structure can impact performance even when two designs have similar
resource usage. One main feature of the \AESmodes{} design is that it operates
on a single column of the AES cipher state, which yields the low resource usage,
but also causes a higher latency. The design also supports the Galois Counter
mode of operation.

\para{\AESpaar{} architecture.} Drimer et al.\ built an area-efficient design
that uses a single-column-based architecture, with four instances of the
datapath that together effectively operate on the full cipher
state~\cite{bib:cmp-paar}. It features an 8 stage inner-pipelined and iterative
structure, and has similar area-efficiency to \sysname{}. However, this design
only supports one process (encryption or decryption depending on build
configuration), whereas \sysname{} supports both encryption and decryption for
12 concurrent blocks. We include it as an example of how differences in the
pipelined part of an iterative design can either increase or decrease
functionality. We also investigate an unrolled variant of this architecture
called \AESexpand{} when evaluating power consumption in
Section~\ref{sec:power}.

\begin{table*}\centering
  \caption{Architecture performance comparison}
  \sffamily
  \ra{1.3}
  \begin{tabular}{@{}lrrrrrcrrrl@{}}
   \toprule 
    & \multicolumn{5}{c}{Resource Usage (\#)}
    & \phantom{abc}
    & \multicolumn{3}{c}{Performance Metrics}\\
    \cmidrule{2-6}
    \cmidrule{8-10}
 
        &\multicolumn{1}{r}{\textbf{Slices}} & 
        \multicolumn{1}{r}{\textbf{LUTs}} & 
        \multicolumn{1}{r}{\textbf{Flip Flops}} & 
        \multicolumn{1}{r}{\textbf{B. RAMs}} & 
        \multicolumn{1}{r}{\textbf{DSPs}} &
        &
        \multicolumn{1}{r}{\textbf{Freq.}} &
        \multicolumn{1}{r}{\textbf{Latency}} &
        \multicolumn{1}{r}{\textbf{Throughput}} &
        \multicolumn{1}{l}{\textbf{Device}}\\
        & & & & & & &
        \multicolumn{1}{r}{(MHz)} &
        \multicolumn{1}{r}{(Cycles)} &
        \multicolumn{1}{r}{(Gbps)} \\
   \midrule 
    \textbf{\AESencdec{}}    & 5613  & 15919 & n/a  & 400   & 144   && 611       & 55       & 78.22      & Virtex 6 \\
    \textbf{\AESmodes{}}     & 159   & n/a   & n/a  & 15    & 28    && 91.5      & 188      & 0.124      & Artix 7 \\
    \textbf{\AESpaar{}}      & 296   & 393   & 665  & 9     & 16    && 550       & 84       & 6.7        & Virtex 5 \\
    \quad Datapath  & 259   & 338   & 624  & 8     & 16    &&   &      &        &  \\
    \textbf{\sysname{}}      & 310   & 909   & 593  & 16    & 18    && 528       & 115      & 7.055      & Zynq 7000 \\
    \quad Datapath & 167   & 266   & 256  & 12    & 18    &&       &     &      &  \\
    \bottomrule
  \end{tabular}
  
  \label{tab:compare}
\end{table*}

\subsection{Architectural Effects on Performance}
\label{sec:performance}

The three comparable designs all use block RAM and DSP in similar ways to
\sysname{} to implement the mix columns and add round key sub-round
transformations. However, the architectural design decisions made in each
implementation heavily influence performance with respect to resource usage.

\AESencdec{} uses a fully-unrolled and pipelined architecture, while the other
designs have an iterative structure with either inner or outer pipeline
registers. \AESencdec{} achieves a throughput of 78.22 Gbps, which is 11 times
the throughput of \sysname{}, however it also uses 17 times the number of LUTs,
25 times the number of block RAM, and 8 times the number of DSP that \sysname{}
uses. Using a fully-unrolled architecture causes an increase in resource usage
that is disproportionate to the performance gain.

Conversely, \AESmodes{} uses an iterative architecture with outer-pipeline
registers that store the block passing through the cipher, and has lower area
than \sysname{}, but also significantly lower throughput. The design uses half
as many logic slices and 1.5 times as many DSP as \sysname{}, but only achieves
2\% of the throughput that \sysname{} supports. Even with similar resource usage
and resource types, the performance of the designs differ greatly. \sysname{}
achieves its higher throughput by using block RAM and DSP to add stages to the
datapath pipeline.

The \AESpaar{} datapath uses a similar iterative inner-pipelined architecture to
\sysname{}, but with 72 more LUTs, 368 more flip flops, and 4 less block RAM. 
This difference in block RAM comes mainly from the design decisions that enable
\sysname{} to perform both encryption and decryption concurrently on 12 blocks
of data. \AESpaar{} uses a T-box-based implementation which combines the sub
bytes and mix columns sub-round transformations into a single look up and
\texttt{XOR} operation~\cite{bib:tech-tbox}. However, this technique utilizes an
entire 36 kilobit block RAM to store the lookup values for encryption, so it is
not possible to perform both encryption and decryption using a single block RAM
configuration.

By keeping the sub bytes and mix columns sub-round transformations separate,
\sysname{} is able to perform both encryption and decryption at the expense of
using 4 more block RAMs. This technique increases the latency of \sysname{} by
two extra delay cycles for the sub bytes sub-round transformation, but also
increases the capacity of the datapath to operate on two more blocks of data
than if the design used T-boxes. Even with extra latency, \sysname{} maintains
higher throughput than \AESpaar{}, and supports both encryption and decryption.

The \sysname{} key schedule was also designed to be able to provide any round
key at any time, in order to support encryption and decryption for 12 concurrent
blocks in the datapath. In order to achieve this, the key schedule utilizes 616
LUTs, 303 flip flops, and 4 block RAMs, which increases the total LUT usage
beyond that of \AESpaar{}. But, at the expense of these extra resources,
\sysname{} achieves more functionality. This shows that including extra
algorithm optimizations such as T-boxes can reduce the potential for design
functionality, and that minimally increasing resource usage can allow for extra
functionality, such as supporting encryption and decryption concurrently on
multiple blocks.

\para{Summary:} Pipelined designs are able to achieve high throughput at the
cost of higher resource usage. Importantly, the increase in resources may
outweigh the increase in throughput. In contrast, slightly reducing resource
usage and pipelining can drastically lower throughput. We conclude that
maximizing the number of pipeline stages in an iterative design without
increasing resource usage results in better area-efficiency. Furthermore, using
algorithmic optimizations may limit functionality such as processing multiple
concurrent blocks and performing both encryption and decryption.  \sysname{}
achieves greater area-efficiency than the other designs by balancing low
resource usage and high-throughput, while offering the advanced functionality
of interleaving encryption and decryption operations on 12 concurrent blocks of
data.

\subsection{Implementation Efficiency}
\label{sec:efficiency}

Studies that build FPGA designs for data processing often evaluate the
efficiency of their implementation by computing the amount of throughput
produced per logic slice. Each of the 3 comparable designs reports their
efficiency in this way, however this metric does not give any information on how
much the block RAM and DSP resources contribute to the design
throughput. For example, \AESencdec{} reports an efficiency of 13.9 Mbps/slice,
although the design also uses 400 block RAM and 144 DSP. Furthermore, this
metric does not incorporate the usage of each logic slice. Since each slice in a
design could be using from 1 LUT and 1 flip flop to 4 LUTs and 8 flip flops, the
throughput per slice metric does not accurately reflect how well the slice logic
is being used to process data.

Starting with this metric, the designs have the following efficiencies:
\AESencdec{} achieves 13.9 Mbps/slice, \AESmodes{} achieves 0.778 Mbps/slice,
\AESpaar{} achieves 22.6 Mbps/slice, and \sysname{} achieves 22.75 Mbps/slice.
Aside from the fact that these calculations do not address the slice utilization
issue previously described, these calculations also neglect whether all of the
slices in a design are part of the datapath, key schedule, or controller. As the
controller and key schedule do not process input data directly, it may not be
appropriate to include their slice usage in such efficiency measurements. This is more
of a philosophical question of which we leave further discussion for future
research.

To address these two issues, we propose that evaluation of design efficiency
should incorporate the following metrics for resources in the datapath only:
Mbps/LUT, Mbps/flip flop, Mbps/block RAM multiplied by the average block RAM
memory usage, and Mbps/DSP.

The LUT and flip flop metrics would be useful in applications where a limited
number of logic slices are available, as it shows how well the implementation
would capitalize on the remaining available logic slice elements. On the other
hand, the block RAM and DSP metrics would be informative in the case where these
resources are limited, and a designer is concerned about whether a design uses
them to their full potential. Additionally, we incorporate the percentage of
memory in the block RAMs that is utilized into the metric to indicate how
effectively the memory is used.

As Drimer et al.\ address in their study, we suggest that future studies report
as much information as possible about their implementations in order to increase
transparency when evaluating AES designs, as we do in this
paper. For example, neither \AESmodes{} nor \AESpaar{} state how their control
mechanisms contribute to resource usage, and neither \AESencdec{} nor
\AESmodes{} state how their key schedules contribute to resource usage. This
makes it difficult to accurately evaluate their efficiency using these metrics.

The adjusted efficiency measurements for \AESpaar{} and \sysname{} is shown in
Table~\ref{tab:efficiency}, as these two designs state the resource usage for
the datapath alone. This table also reveals that it is effective to use block
RAM and DSP in the AES datapath, as each singular unit is able to process 
large chunks of data (32-bit for block RAM, 48-bit for DSP), instead of
splitting operations across multiple LUTs and flip flops.

\begin{table} \centering
 \caption{Throughput Per Resource (Mbps/\#)}  
  \sffamily
  \ra{1.3}
  \begin{tabular}{@{}lrrrr@{}}
    \toprule
& \textbf{LUT}    & \textbf{Flip Flop} & \textbf{B. RAM}
                    & \textbf{DSP}    \\
    \midrule 
    \textbf{\AESpaar{}}      & 19.8   & 10.7      & 837.5     & 418.75 \\    
    \textbf{\sysname{}}      & 26.5   & 27.56     & 220.47    & 391.94 \\    
    \bottomrule
  \end{tabular}
  
  \label{tab:efficiency}
\end{table}

\para{Summary:} The traditional technique of calculating throughput per logic
slice is a weak metric for evaluating design efficiency in terms of performance
versus area. We propose that efficiency analysis include four measurements:
throughput per LUT, throughput per flip flop, throughput per block RAM, and
throughput per DSP. The block RAM measurement should incorporate what
percentage of the block RAM memory is actually used to process data. The DSP
measurement reveals that using block RAM and DSP in the datapath is an
effective way to process large chunks of data. \sysname{} has nearly two times
better LUT and flip flop efficiency than the other area-efficient design in our
evaluation, \AESpaar{}, and similar DSP efficiency with more functionality.

\subsection{Design Effects on Power Consumption}
\label{sec:power}

While there are fewer FPGA AES studies that focus on low-power implementations,
power consumption is still an area of interest for many system designers.
Of the three comparable designs, only the designers of \AESpaar{} report power
consumption. Additionally, Drimer et al.\ include a second fully-unrolled
implementation built on the base structure of the \AESpaar{} design, which we
refer to as \AESexpand{}~\cite{bib:cmp-paar}. The power consumption of these two
implementations and \sysname{} is shown in Table ~\ref{tab:compare-power}.

\begin{table} \centering
  \caption{Datapath power consumption (mW)}  
  \sffamily
  \ra{1.3}
  \begin{tabular}{@{}lrrrrrr@{}}
    \toprule
& \textbf{Logic}   & \textbf{B.RAM}  & \textbf{DSP}    &
         \textbf{Signal} & \textbf{Total} \\
        & \textbf{Slices}   &   &     &
         \textbf{+Clock} &  \\
    \midrule 
    \textbf{\AESpaar{}}        & 56   & 285   & 111    & 39  & 491  \\
    \textbf{\AESexpand{}}      & 165  & 2140  & 833    & 74  & 3212  \\
    \textbf{\sysname{}}        & 7    & 259   & 58     & 88  & 412  \\
    \bottomrule
  \end{tabular}
  \label{tab:compare-power}
\end{table}

\AESpaar{} uses more LUTs and flip flops than \sysname{}, which is the cause of
the higher slice power consumption.  However, \AESpaar{} uses slightly less
block RAM and DSP resources than \sysname{}, which is not reflected in the
power comparison. This may be due to differences in the target device, as
\AESpaar{} is implemented on a Virtex FPGA, which is a high-performance FPGA
that naturally consumes more power than the Zynq 7000 SoCs. This highlights the
importance of disclosing the target device as part of an AES implementation
analysis.

On the other hand, \AESexpand{} consumes nearly 10 times the power of
\sysname{} and \AESpaar{}. With a fully-pipelined structure, this design also
uses significantly more resources. Table~\ref{tab:compare-power} shows that the
large increase in power consumption from \AESpaar{} to \AESexpand{} comes from
the block RAMs and DSP slices, which both have increase factors of 7.5, while
the logic slice power only increases by a factor of 3. This shows that using
more block RAMs and DSP slices in a design will also significantly increase the
power usage. Therefore, while low-area designs certainly consume less power due
to their overall lower resource usage, low-area designs that use block RAMs and
DSP slices will have higher power consumption than equivalent designs that
primarily use logic slices.

Finally, while the power consumption of a running implementation is important
for power supply considerations, it is also desirable to know how much power
over time is required to process data. We propose that an additional power
metric of the nanowatt-seconds required to process a single block is needed to
better measure the tradeoffs between latency and pipeline length. Overall, this
metric reflects how effectively the design uses its power to process data.
The \sysname{} design uses 7.47 nanowatt-seconds, \AESpaar{} uses 9.37
nanowatt-seconds, and \AESexpand{} uses 622.17 nanowatt-seconds to process a
single block of input data.

\para{Summary:} Replacing logic slices with higher-power components, like block
RAMs and DSP slices, will increase the overall power usage, but with careful
design the performance growth of the datapath outpaces this power increase.
Furthermore, the target device directly impacts power consumption, so it is
important for future studies to describe the exact evaluation platform for
accurate power evaluation. Further, we recommend that studies include an
evaluation of their power usage that shows how much power, in nanowatt-seconds,
is required to process a single block of data. This metric better measures the
relationship between power consumption and performance, showing that \sysname{}
consumes less power over time to process a single block of data than \AESpaar{}.

\subsection{Overall Summary}

Area-efficiency requires striking a balance between throughput and resource
usage, primarily by capitalizing on under-utilized components. \sysname{}
strikes this balance by leveraging block RAM tiles and DSP slices key components
in the architecture's compact datapath structure. Further, evaluating
area-efficiency requires the use of new metrics that better encapsulate LUT,
flip flop, block RAM, and DSP usage. These new metrics offer more insight into
how effectively a design uses the resources available in the target device.
\sysname{}'s use of block RAMs and DSP slices  results in an implementation with
similar efficiency to \AESpaar{}, but higher throughput, less power consumption
over time, and greater functionality, i.e., support for multiplexed encryption
and decryption on 12 concurrent blocks.
 
\section{Accelerators on Embedded Systems}
\label{sec:accelerators}

One promising application of area-efficient AES architectures is enabling
efficient cryptographic primitives on resource-constrained embedded systems
without forcing the system designer to choose between using the FPGA for
security or some other operation. In this section, we discuss the feasibility
of using several AES implementations, including \sysname{}, alongside other
hardware accelerators. To do this, we compute how many logic slices, block
RAMs, and DSPs are available after co-location on Zynq 7000 devices.
Complicating our analysis, we can only estimate the number of logic slices
in each accelerator, as the studies only report their LUT usage. 
Therefore, we present a best-case analysis, underestimating the number of logic
slices. Further, the AES designs may run slower after co-location due to
architectural differences between the original devices and the Zynq devices.
We consider the following hardware accelerators for co-location with the AES
designs:

\begin{enumerate}
  \item \textbf{Video Processing:} Hoozemans et al.\ designed a video processing
    system using softcore processors for dedicated image filtering. The system
    utilizes 8,315 slices, 105 block RAMs, and 26 DSPs, which was implemented on
    a Zynq 7000 (xc7z020) SoC. This platform has the following available
    resources: 13,300 slices, 140 block RAMs, and 220 DSPs~\cite{bib:app-video}.

  \item \textbf{DLAU} Wang et al.\ built a deep learning accelerator on a Zynq
    7000 (xc7z020) SoC using 9,096 slices, 35 block RAMs, and 167
    DSPs~\cite{bib:app-dlau}.

  \item \textbf{CNN} Qiu et al.\ built a convolutional neural network
    accelerator for image classification using 45,654 slices, 486 block RAMs,
    and 780 DSPs~\cite{bib:app-cnn}. Their study used a Xilinx ZC706 evaluation
    board which features a Zynq 7000 (xc7z045) with 54,650 slices, 545 block
    RAMs, and 900 DSPs.

  \item \textbf{DNN} Hao et al.\ proposed a methodology for designing
    FPGA-based deep neural network accelerators and presented several different
    network model implementations on a Zynq 7000 (xc7z020)
    SoC~\cite{bib:app-dnn}. We selected three of their new models:
    \textbf{DNN 1} uses 10,973 slices, 134 block RAM, 202 DSPs; \textbf{DNN 2}
    uses 10,161 slices, 109 block RAMs, 186 DSPs; \textbf{DNN 3} uses 11,704
    slices, 108 block RAMs, and 172 DSPs.
    
\end{enumerate}

\begin{table*}\centering
  \caption{Available resources after co-location}
  \sffamily
  \ra{1.3}
  \begin{tabular}{@{}lrrrcrrrcrrrcrrr@{}}
    \toprule
    & \multicolumn{3}{c}{\textbf{\AESencdec{}}}
    & 
    & \multicolumn{3}{c}{\textbf{\AESmodes{}}}
    & 
    & \multicolumn{3}{c}{\textbf{\AESpaar{}}}
    & 
    & \multicolumn{3}{c}{\textbf{\sysname{}}}\\
    \cmidrule{2-4}
    \cmidrule{6-8}
    \cmidrule{10-12}
    \cmidrule{14-16}
        &\multicolumn{1}{c}{Slices} & 
        \multicolumn{1}{c}{B. RAMs} & 
        \multicolumn{1}{c}{DSPs} &
        &
        \multicolumn{1}{c}{Slices} & 
        \multicolumn{1}{c}{B. RAMs} & 
        \multicolumn{1}{c}{DSPs} &
        &
        \multicolumn{1}{c}{Slices} & 
        \multicolumn{1}{c}{B. RAMs} & 
        \multicolumn{1}{c}{DSPs} &
        &
        \multicolumn{1}{c}{Slices} & 
        \multicolumn{1}{c}{B. RAMs} & 
        \multicolumn{1}{c}{DSPs} \\
    \midrule
    \textbf{Video}            & -628 & -365 & 50 && 4,826 & 20 & 166 && 4,689 & 26 & 178 && 4,675 & 19 & 176 \\
    \textbf{DLAU}             & -1,409 & -295 & -91 && 4,045 & 90 & 25 && 3,908 & 96 & 37 && 3,894 & 89 & 25 \\
    \textbf{CNN}              & 3,383 & -341 & -24 && 8,837 & 31 & 92 && 8,700 & 50 & 104 && 8,686 & 43 & 102 \\
    \textbf{DNN 1}            & -3,286 & -394 & -126 && 2,168 & -9 & -10 && 2,031 & -3 & 2 && 2,017 & -10 & 0 \\
    \textbf{DNN 2}            & -2,474 & -369 & -110 && 2,980 & 3 & 6 && 2,843 & 22 & 18 && 2,829 & 15 & 16 \\
    \textbf{DNN 3}            & -4,017 & -368 & -96 && 1,437 & 17 & 20 && 1,300 & 23 & 32 && 1,286 & 16 & 30 \\
    \bottomrule
  \end{tabular}
  \label{tab:cases}
\end{table*}

Table~\ref{tab:cases} shows the amount of remaining resources after co-locating
each AES design with each non-AES accelerator---a negative resource value in any
column indicates that the two accelerators cannot fit on the same FPGA.
\AESencdec{} does not fit alongside any of the accelerators due to the high
resource usage of its fully-unrolled architecture. While this design produces
high throughput, it is not suitable for co-location with other accelerators.
\AESmodes{}, \AESpaar{}, and \sysname{} all fail when co-located with DNN 1,
due to the DNN accelerator's high block RAM and DSP usage. However, in all
other cases \sysname{} and the other two AES designs are small enough to be
implemented alongside the other accelerators. Among these three, \sysname{} is
the most attractive option as it offers the most throughput and supports
concurrent encryption and decryption. 

\para{Summary:} Area-efficiency is important for allowing co-location of
accelerators on the same FPGA. \sysname{} makes the most efficient use of the 
available resources to support the most functionality and produce the highest
throughput among the AES designs that support co-location. \section{Related Work}
\label{sec:related-work}

In contrast to \sysname{} and the architectures discussed previously, there are
a number of proposed architectures that do not leverage block RAMs or DSPs, or
do not provide detailed explanations of how they use block RAMs and DSPs. Few of
these architectures are designed for area-efficiency, but are included to
provide a broader context for our work. Below, we broadly divide these into the
following categories based on their structure: (1) instruction-based, (2)
purely iterative with no pipeline, (3) iterative with inner pipelining, and (4)
fully pipelined and unrolled.

Norbert et al.\ proposed an uncommon instruction-based design, where an external
interface controls data through the cipher. This design produces 215 Mbps of
encrypted (or decrypted) data running on a 161 MHz clock, and utilizes 1,125
logic slices~\cite{bib:inst0}. Tay et al.\ proposed a design that has a purely
iterative architecture which produces 597 Mbps of encrypted data running on a
107 MHz clock, and utilizes 3,048 LUTs and 808 flip flops~\cite{bib:iter0}.

Rahmunnisa et al.\ proposed an iterative inner-pipelined design that produces
37.1 Gbps of decrypted data running on a 505 MHz clock, and utilizes 3,788 LUTs,
2,056 flip flops, 48 block RAMs, and 2 DSP slices~\cite{bib:ipit0}. Although
this design uses block RAM and DSP, the nature of their use was not explicitly
described, so we did not include the design in our detailed analysis. Farashahi
et al.\ proposed an iterative inner-pipelined design that produces 7.95 Gbps of
encrypted data running on a 671.5 MHz clock, and utilizes 3,557 LUTs, and 2,132
flip flops~\cite{bib:ipit1}. Rao et al.\ proposed an iterative inner-pipelined
design that produces 676 Mbps of encrypted data running on a 311.7 MHz clock,
and utilizes 359 logic slices~\cite{bib:ipit2}.

Zhang et al.\ proposed a fully-unrolled and pipelined design that produces 93.5
Gbps of encrypted data running on a 730.7 MHz clock, and utilizes 5,081 logic
slices~\cite{bib:unrl0}. Samiee et al.\ proposed a fully-unrolled and pipelined
design that produces 43.71 Gbps of decrypted data running on a 341.5 MHz clock,
and utilizes 7,865 logic slices~\cite{bib:unrl1}. Oukili and Bri proposed a
fully-unrolled and pipelined design that produces 79 Gbps of encrypted data
running on a 617.6 MHz clock, and utilizes 14,736 LUTs and 18,305 flip
flops~\cite{bib:unrl2}.

While these related works produce high throughput, it may be
difficult to use the nearby block RAMs and DSPs due to logic congestion and
routing challenges. Comparatively, the \sysname{} architecture is significantly
more area-efficient due to its use of block RAMs and DSPs. By offloading some
processing from logic slices, \AESpaar{} and \sysname{} make better use of 
available resources. Furthermore, \sysname{} achieves similar area-efficiency as
\AESpaar{}, but offers more functionality. \section{Conclusions}
\label{sec:conclusion}

While AES and other cryptographic primitives form the foundation of security,
there are myriad safety-critical embedded systems that lack the ability to do
these operations efficiently. Consequently, such systems often forgo
cryptographic operations in favor of saving their limited computational and
power resources for other tasks. 

FPGA-based hardware acceleration offers a performant, low-power solution to
cryptography on embedded systems. Just as importantly, FPGAs also offer the
flexibility to \emph{concurrently} accelerate other emerging operations, e.g.,
deep learning. However, taking advantage of these new possibilities requires a
new approach to developing accelerator architectures, namely one that focuses
on balancing raw performance with resource usage so that system designers do not
have to choose between using the FPGA for cryptographic operations or using it
for other algorithms. In short, area-efficient designs make better use of
available resources to allow for co-location of hardware accelerators.

We proposed an area-efficient AES architecture, \sysname{}, that achieves a
balance between resource usage and throughput by incorporating under-utilized
components, such as block RAM tiles and DSP slices. We identified how
architectural design decisions influence key trade-offs and  discussed new
metrics for evaluating the efficiency and power usage of cryptographic
accelerators. These metrics provide a measure of how effectively a design
capitalizes on the available resources to improve performance. \sysname{}
achieves higher resource efficiency than throughput-focused AES architectures,
higher throughput than low-area designs, and more functionality and lower power
usage than other area-efficient designs. While our focus is on AES, the design
principles we discuss in this paper are applicable to other cryptographic
algorithms.

Area-efficient hardware accelerators promote the adoption of security for new
and emerging applications. Such designs, for example, may allow robotic swarm
systems with limited resources to offload parts of network security operations,
such as TLS, to hardware. Regardless of the specific application, area-efficient
hardware acceleration is a promising approach for boosting the design of secure
embedded systems. 
\bibliographystyle{ACM-Reference-Format}
\bibliography{bib}

\appendix
\clearpage{}\section{Xilinx 7 Series FPGAs}
\label{apdx:fpga}

\begin{figure}[tb]
  \centering
  \includegraphics[width=0.5\textwidth,keepaspectratio]{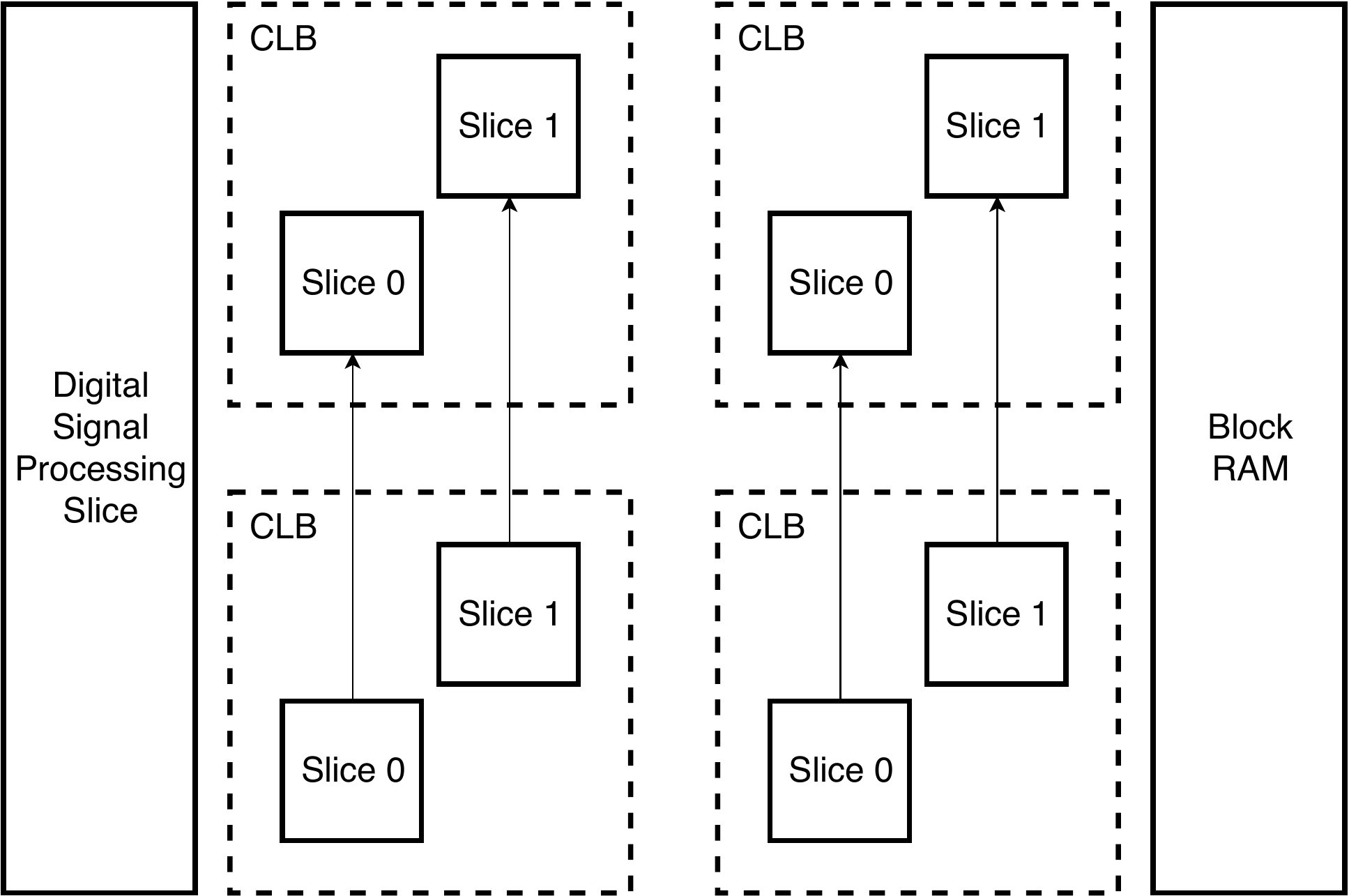}
  \caption{Simplified Xilinx 7-Series FPGA Architecture}
  \label{fig:clb-simp}
\end{figure}

\begin{figure}[tb]
  \centering
  \includegraphics[width=0.5\textwidth,keepaspectratio]{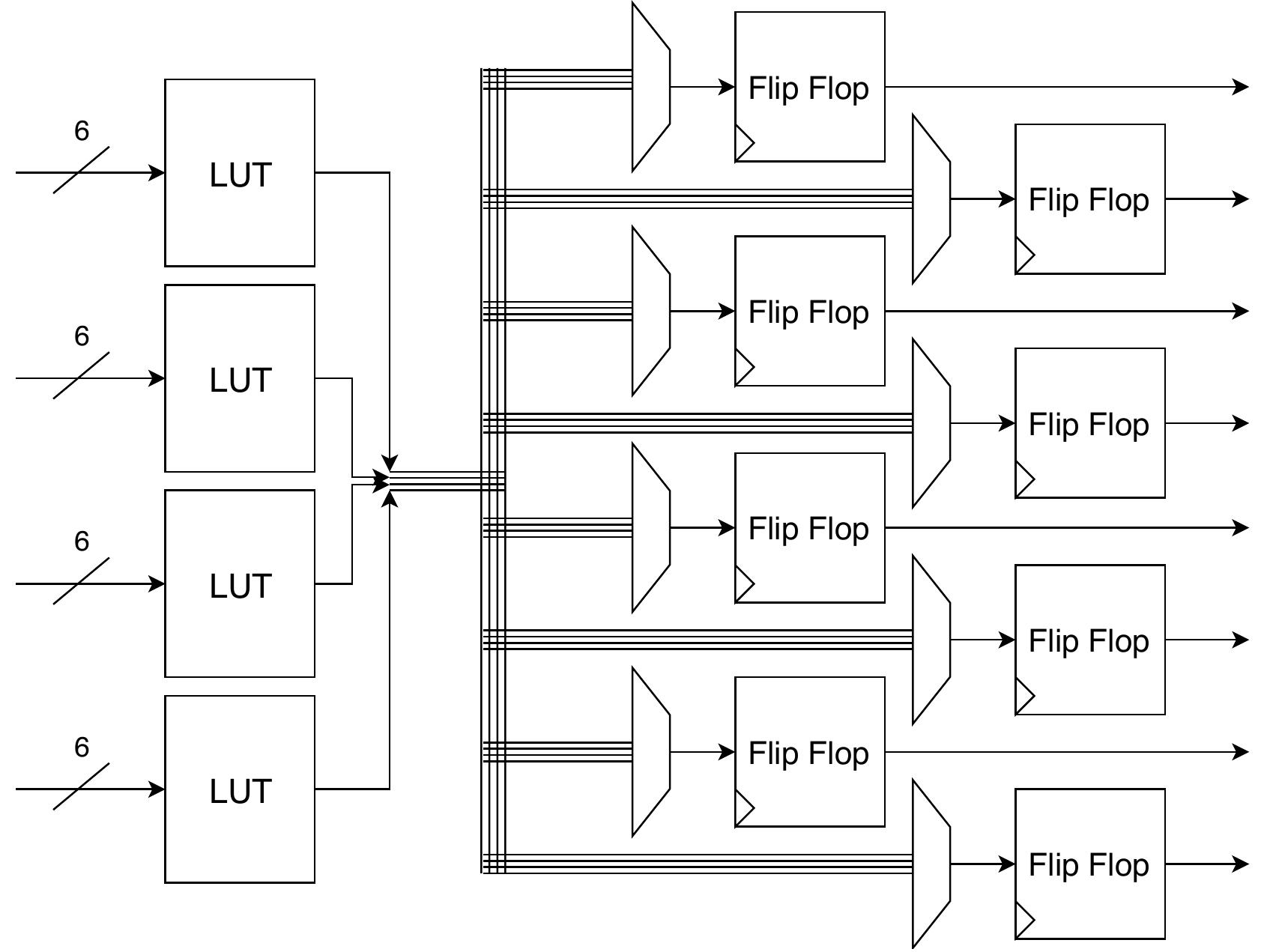}
  \caption{Simplified Xilinx 7-Series Logic Slice}
  \label{fig:slice-simp}
\end{figure}

FPGAs have a column-based structure, where groups of configurable logic blocks
lie in between columns of block RAMs and DSPs, as shown in
Figure~\ref{fig:clb-simp}. Within each logic block there are two logic slices
that contain 4 LUTs and 8 flip flops (Figure~\ref{fig:slice-simp}). We argue
that area-efficient designs are needed to take advantage of available resources
due to to tight packing of components in the FPGA. A design that only uses the
slices, or just the block RAM and DSP, in an area could make it difficult for
others to access the unused resources. By using all of the available resources
in a given area, a design can increase its performance and reduce resource
waste.

\begin{figure}[tb]
  \centering
  \includegraphics[width=0.25\textwidth,keepaspectratio]{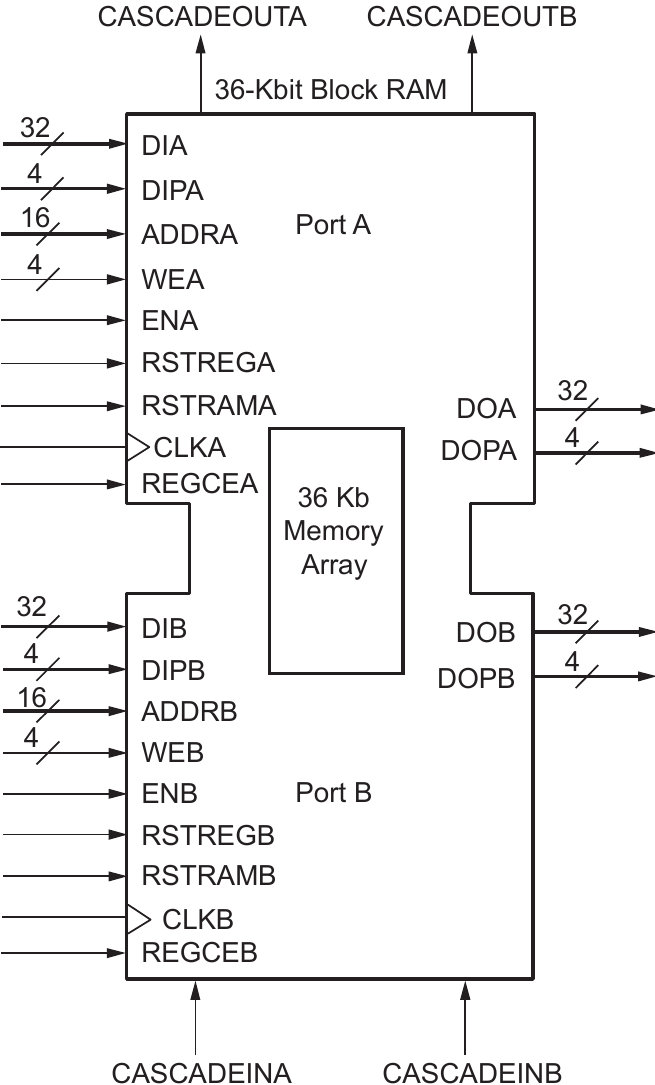}
  \caption{Xilinx 7-Series Dual Port Block RAM}
  \label{fig:bram}
\end{figure}

The Xilinx 7-series block RAM supports two independent data ports that access
the same bank of memory as shown in Figure~\ref{fig:bram}. The dual port feature
is helpful in a wide range of applications, and we take advantage of this by
using a single block RAM to perform two byte lookups in the sub bytes and mix
columns transformations. Furthermore, the input address and output data size are
flexible, which also allows different use cases where data sizes may not match
for each block RAM. 

\begin{figure}[tb]
  \centering
  \includegraphics[width=0.5\textwidth,keepaspectratio]{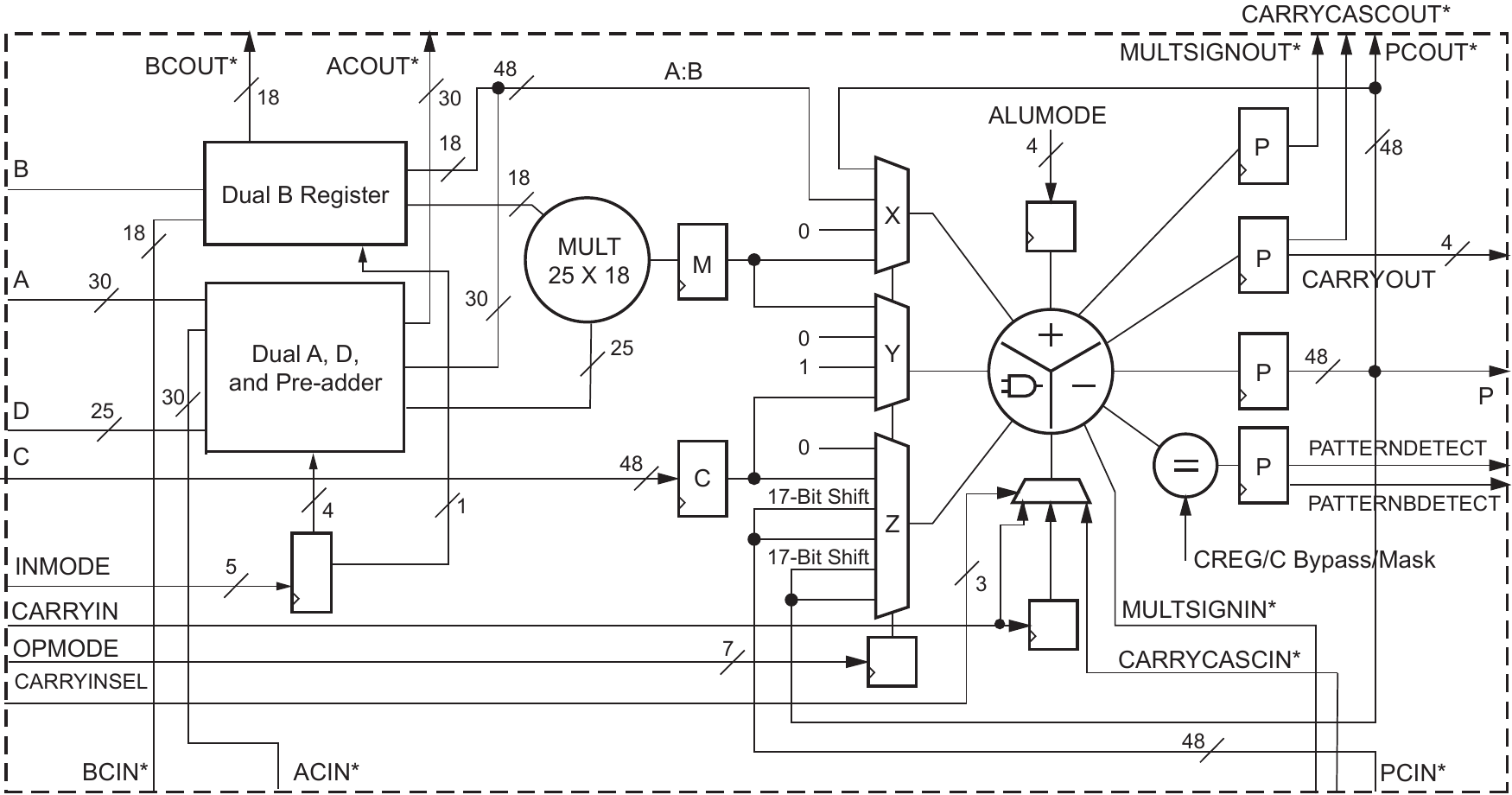}
  \caption{Xilinx 7-Series DSP48E1 Slice}
  \label{fig:dsp}
\end{figure}

The Xilinx 7-series DSP slices support a number of arithmetic and logic
functions, which allows them to be used in a number of applications, such as
AES, other than signal processing. The logic functions operate on two inputs,
which have optional input registers, as shown in Figure~\ref{fig:dsp}, to
decrease the critical path from other components. Additionally, DSP slices have
dedicated inputs and outputs connected to other vertically-aligned DSPs for
high-speed cascaded calculations.

\clearpage{}
\clearpage{}\section{Mix Columns Matrix Equations}
\label{apdx:mc}

\begin{figure}[b]
  \centering
  \includegraphics[width=0.3\textwidth,keepaspectratio]{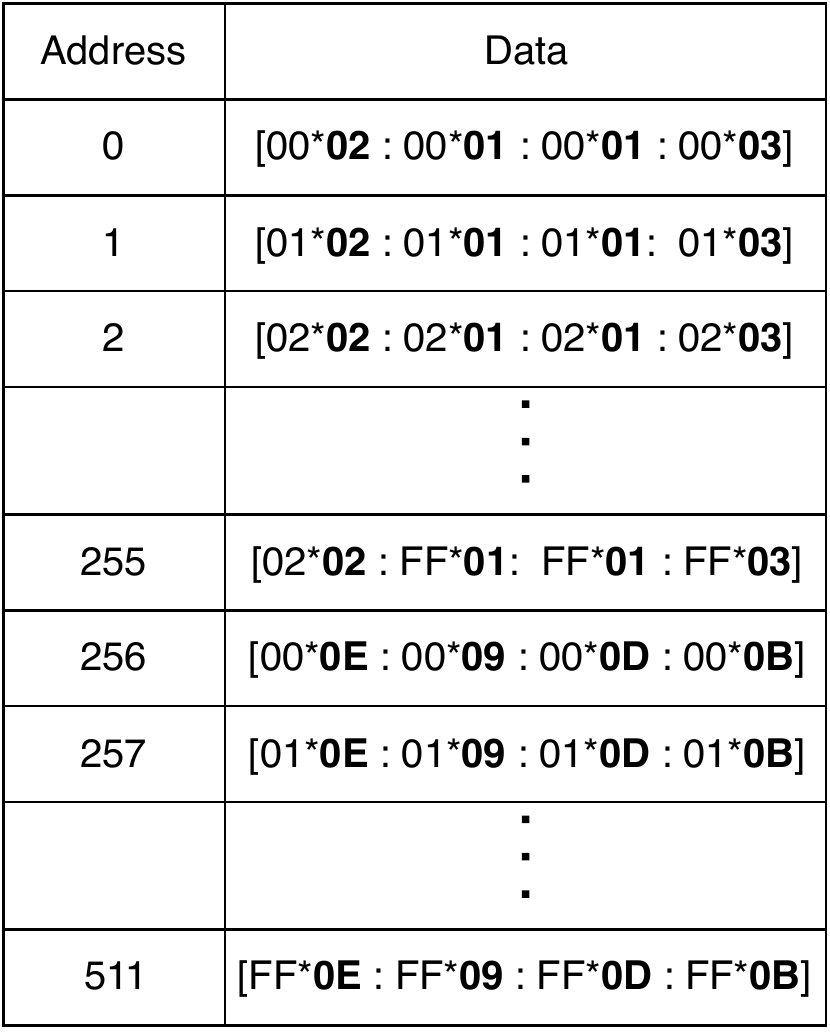}
  \caption{Mix Columns Multiplication Storage}
  \label{fig:mc-bram-layout}
\end{figure}

The mix columns transformation performs matrix multiplication between the cipher
state block and a fixed encryption or decryption matrix, shown below. The
multiplication involves each byte of the cipher state block, and each byte of
the fixed matrix, so we store the products of each input byte as a 32-bit vector
in block RAM. Since the matrices for encryption and decryption are congruent,
but with different values, we use the same storage format order for encryption
and decryption, as shown in Figure~\ref{fig:mc-bram-layout}. 

\[
  \stackrel{\mbox{Encryption Matrix}}{\begin{bmatrix}
    02 & 03 & 01 & 01 \\
    01 & 02 & 03 & 01 \\
    01 & 01 & 02 & 03 \\
    03 & 01 & 01 & 02
    \end{bmatrix}}\ \quad
  \stackrel{\mbox{Decryption Matrix}}{\begin{bmatrix}
    0E & 0B & 0D & 09 \\
    09 & 0E & 0B & 0D \\
    0D & 09 & 0E & 0B \\
    0B & 0D & 09 & 0E
    \end{bmatrix}}
\]\\

We use the following equations (\ref{eq:mc-vectors}) for both encryption and
decryption to form the input vectors for the DSP \texttt{XOR} phase of mix
columns (see Section~\ref{sec:mc}). These equations are for the high bits of the
output block, and we use similar patterns for the middle and low bits.

\begin{gather}
  \begin{split}
    Vec_0 = RAM[s_{0,0}][31:24] + RAM[s_{0,0}][23:16] + \\
        RAM[s_{0,0}][15:8] + RAM[s_{0,0}][7:0] + \\
        RAM[s_{0,1}][31:24] + RAM[s_{0,1}][23:16] \\\\
    Vec_1 = RAM[s_{1,0}][7:0] + RAM[s_{1,0}][31:24] + \\
        RAM[s_{1,0}][23:16] + RAM[s_{1,0}][15:8] + \\
        RAM[s_{1,1}][7:0] + RAM[s_{1,1}][31:24] \\\\
    Vec_2 = RAM[s_{2,0}][15:8] + RAM[s_{2,0}][7:0] + \\
        RAM[s_{2,0}][31:24] + RAM[s_{2,0}][23:16] + \\
        RAM[s_{2,1}][15:8] + RAM[s_{2,1}][7:0] \\\\
    Vec_3 = RAM[s_{3,0}][23:16] + RAM[s_{3,0}][15:8] + \\
        RAM[s_{3,0}][7:0] + RAM[s_{3,0}][31:24] + \\
        RAM[s_{3,1}][23:16] + RAM[s_{3,1}][15:8]
  \end{split}
\label{eq:mc-vectors}
\end{gather}
\\
\clearpage{}
\clearpage{}\section{Physical Layout of \sysname{}}
\label{apdx:physical}

\begin{figure}[tb]
  \centering
  \includegraphics[width=0.5\textwidth,keepaspectratio]{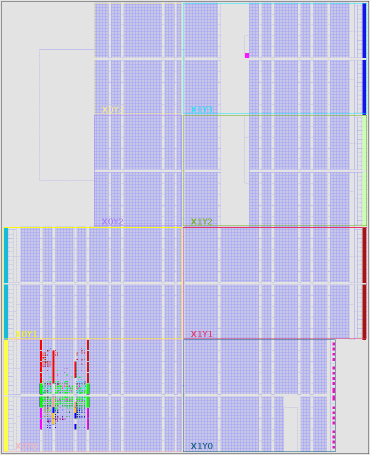}
  \caption{Location of \sysname{} in the Virtex 7 FPGA}
  \label{fig:device}
\end{figure}

We implemented the \sysname{} architecture on a Zynq 7000 (xc7z030) SoC.
Figure~\ref{fig:device} shows that the implementation fits in one half of a
clock region, which can save power in applications where the other clock regions
are not used. Since the design is compact, the device does not need to use extra
power to route clock and control signals to multiple clock regions.

\begin{figure}[tb]
  \centering
  \includegraphics[width=0.5\textwidth,keepaspectratio]{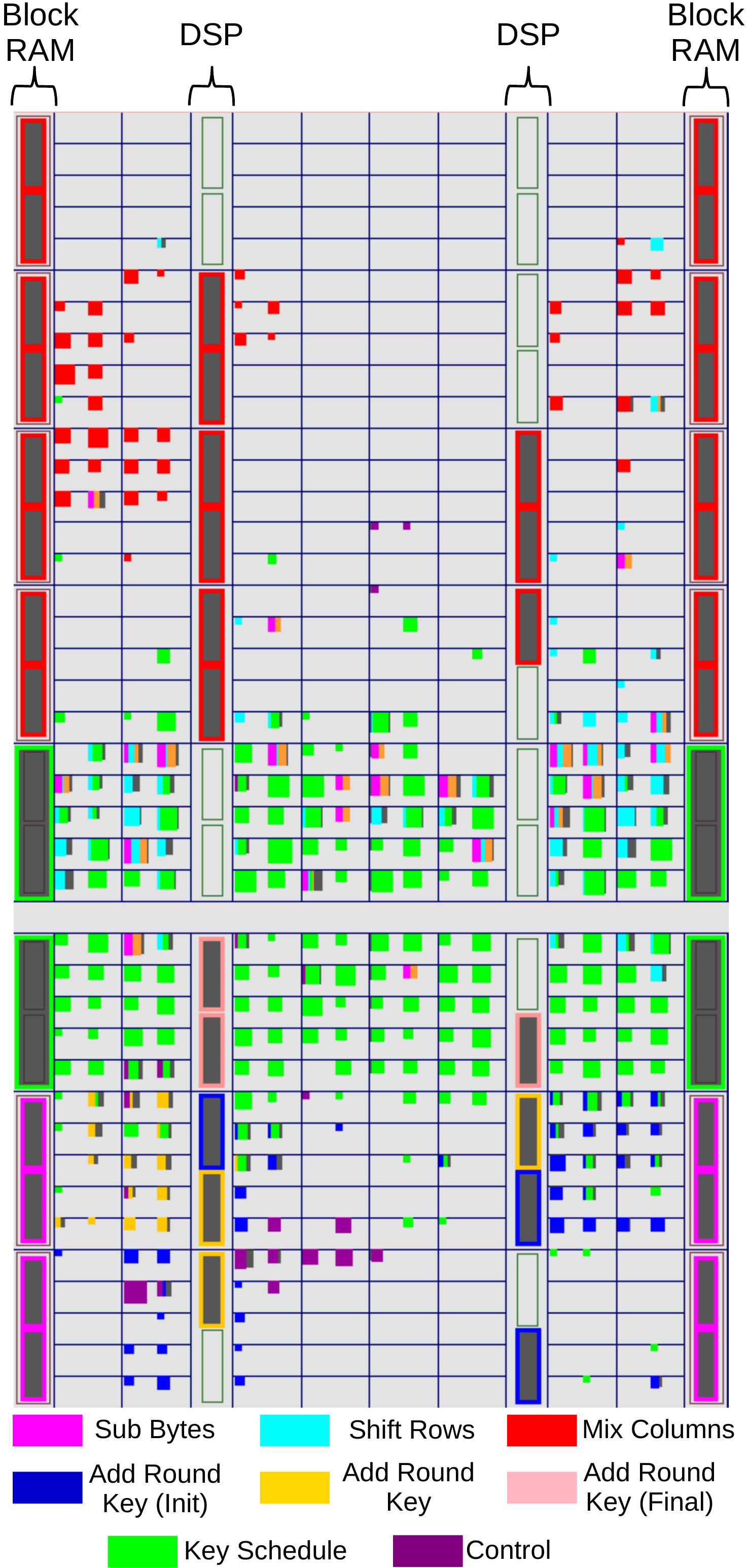}
  \caption{Physical Layout of \sysname{}}
  \label{fig:core}
\end{figure}

Figure~\ref{fig:core} shows that the \sysname{} implementation lies between two
columns of block RAMs, and uses most of the available DSPs between them.
Furthermore, the larger filled-in squares between the block RAMs and DSPs
indicate high usage of the LUTs and flip flops available in the logic slices.
This shows that \sysname{} makes effective use of nearby resources, and prevents
slices, block RAMs, and DSPs from being unused. If \sysname{} did not use the
available block RAMs and DSPs, it would be difficult for other co-located
accelerators to use them due to the occupied slices surrounding them.

\clearpage{}
\end{document}